\documentclass[sigconf]{acmart}

\copyrightyear{2024}
\acmYear{2024}
\setcopyright{rightsretained}
\acmConference[RecSys '24]{18th ACM Conference on Recommender Systems}{October 14--18, 2024}{Bari, Italy}
\acmBooktitle{18th ACM Conference on Recommender Systems (RecSys '24), October 14--18, 2024, Bari, Italy}
\acmDOI{10.1145/3640457.3688060}
\acmISBN{979-8-4007-0505-2/24/10}

\begin{document}

\title{AI-assisted Coding with Cody: Lessons from Context Retrieval and Evaluation for Code Recommendations}


\author{Jan Hartman, Rishabh Mehrotra, Hitesh Sagtani, Dominic Cooney, Rafal Gajdulewicz, Beyang Liu, Julie Tibshirani, Quinn Slack}
\email{{jan.hartman,rishabh.mehrotra}@sourcegraph.com}



\affiliation{%
  \institution{Sourcegraph}
  \city{San Francisco}
  \state{California}
  \country{USA}
}

\begin{abstract}
In this work, we discuss a recently popular type of recommender system: an LLM-based coding assistant. Connecting the task of providing code recommendations in multiple formats to traditional RecSys challenges, we outline several similarities and differences due to domain specifics. We emphasize the importance of providing relevant context to an LLM for this use case and discuss lessons learned from context enhancements \& offline and online evaluation of such AI-assisted coding systems.

\end{abstract}



\maketitle

\section{Introduction}

Recent advancements in the large language model (LLM) area have made it feasible to use such models to assist with software development. Popular models like the GPT family~\cite{achiam2023gpt} or Claude~\cite{anthropic2024claude} have demonstrated good results on coding benchmarks such as HumanEval~\cite{chen2021evaluating}. There is also a number of LLMs fine-tuned for code generation~\cite{chen2021evaluating} 
with many such models having publicly available weights~\cite{li2023starcoder,roziere2023code}. These models are commonly used in coding assistant applications to improve productivity~\cite{fan2023} by providing code recommendations in use cases such as autocomplete or chat~\cite{ross2023}.

A key part of any coding assistant is understanding the user's codebase as this enables the assistant to provide reliable recommendations. However, a key problem here is that the underlying LLMs likely are not trained on the user's code (typically due to costs, UX issues, and privacy concerns), meaning that out of the box, their responses will be far from optimal. Furthermore, lacking repository-specific code context, the generated code often suffers from hallucinations.
The common solution to this problem is providing relevant context to the model inside the prompt.
This involves finding sections of text or code that are relevant to the user's query and stitching together a prompt to enable in-context learning~\cite{dong2022survey} for the LLM, dramatically increasing the quality of the responses~\cite{brown2020language}.

However, we cannot simply fit all relevant snippets into the prompt as the context window size in LLMs is limited for practical reasons. Despite many recent advancements~\cite{chen2023extending} in this direction, some even allowing sizes up to 1 million tokens~\cite{team2023gemini}, extending the context window is not a one-size-fits-all solution -- it makes inference far more costly and with large codebases, we still might not be able to fit all relevant items. This means that our approach has to consider both recall and precision and be able to pick only a few most relevant context items.

In the remainder of this work, we discuss the importance of the context-picking engine inside a coding assistant, outline some key evaluation differences compared to a traditional RecSys, and provide some practical considerations gathered while working on a real-life coding assistant -- Sourcegraph's Cody\footnote{\url{https://cody.dev}}.



\section{Context Engine}

The goal of the context engine inside a coding assistant is similar to candidate generation in a traditional recommender system: given the state of the user's workspace and a query, find N most relevant sections of text or code (we call these context items). The number of context items depends heavily on the use case as the latency can be a critical part of the recommendations -- an autocomplete suggestion needs to be generated much faster than a chat response. Like in most RecSys, we also distinguish between two stages in this process: retrieving context items and ranking them. 


The number of sources of context we can use in a coding assistant is vast: local and remote code, source control history, code review tools, editor state, terminal, documentation, chats, internal Wikis, ticketing systems, observability dashboards, etc. 
Unlike a traditional RecSys, items from these are not in a centralized source of data and we often do not have access to them ahead of time, thus diminishing our ability to e.g. create indexes or precompute useful metrics.

For searching through the space of context items, we can use techniques like similarity-based matching, keyword search, semantic search (embedding-powered retrieval augmented generation ~\cite{lewis2020retrieval}), code graph analysis, etc. In this stage, approximate approaches are preferred for performance reasons and we are optimizing for recall -- that is, we care more about retrieving all relevant items rather than retrieving only relevant items. Our limitations here are latency and the token budget.
An interesting observation is that context retrieval sources should be complementary -- that is, we
want them to retrieve distinct sets of relevant items. Theoretically, we should be able to achieve this by utilizing techniques like keyword and semantic search as they use different matching strategies.

The second stage of the context engine consists of ranking the context items. Here, our use case again differs from traditional RecSys: we do not present the items we are ranking to the user, the goal of ranking here is only to pick the most relevant items to insert into the LLM prompt. 
This means we cannot get user feedback as labels, which complicates evaluation -- we discuss more in the next section. Furthermore, we do not take into account the positions of the ranked items. We only want to pick items above a certain threshold. The threshold can be e.g. a relevance score or a token budget. Given a good ranking model, this stage optimizes for precision, i.e. keeping only relevant items.
With our setting, we can make a pointwise ranker model work well: we can simply train it to predict whether a given context item is relevant to the user's query. 


\section{Evaluation}

Evaluating an LLM-based coding assistant can present several difficulties. The biggest issue is the large online-offline discrepancy as the two settings are significantly different. 
It is important to note that we cannot rely on being able to easily store the complete state of the workspace at request time due to several factors -- most prominently, technological issues and privacy concerns. Many of the context item sources also only exist on the user's device and can be ephemeral. Being unable to log and use this data for offline evaluations makes it very difficult to bridge the online-offline gap. In the online environment, we can receive user feedback -- we know if the user accepted an autocomplete suggestion or if they marked the chat response as useful / not useful. These can be used to judge online A/B tests.

The second issue is a significant lack of labeled data, especially with regard to the context engine. A dataset here might be a set of queries against codebases with the labels being sets of relevant snippets of text or code. Even manual annotation is no easy task: to get correct answers to queries like ``Where is XYZ implemented?'' in real codebases, we require an expert annotator, meaning that this task cannot be easily crowdsourced. 

Lastly, we have to take into account that the coding assistant has several components -- we can try to evaluate either each component (retrieval, ranking, LLM) separately or everything end-to-end. 
For retrieval, we have created an internal, crowdsourced dataset of queries and have found that can provide great insights into different retrieval strategies.

To build and evaluate the model used for ranking, we share learnings from various datasets we created that enabled us to perform offline evaluations of various coding assistant features: code completions, code edits, unit test generation, and open-ended chat. 

The end-to-end case is by far the closest to the online setting. 
Of course, the main goal of the coding assistant is to provide high-quality recommendations. End-to-end evaluation means we have to judge the utility to the user, which differs based on the task.
Here, one of the advantages of working with code is that we can utilize domain specifics to make sure that the recommendations provide value. For autocompletions, this can involve syntactic (e.g. does it parse?) and semantic (e.g. do the types match?) checks over the resulting code segment to prevent nonsensical outputs. 
When generating unit tests, we can follow similar logic, including checking if the tested function was called, if it was called with the correct arguments, and actually running the test. If the user poses a question relating to existing functionality in the codebase, we can scan the resulting code segment to make sure the generated code symbols exist.
For general chat responses, we can utilize techniques such as LLM judging~\cite{zheng2024judging} to evaluate their quality.
An especially positive side of these checks is that certain ones can be used as guardrails in the online setting: if the assistant generates a recommendation, it makes sense to check if it fulfills the intended purpose before displaying it to the user. With techniques such as these, we can curtail LLM hallucinations and improve the quality of the assistant.






\section*{Presenter bio}
Jan Hartman is a machine learning engineer at Sourcegraph, where he works on Cody's context engine. His work is focused on evaluating new sources of context, i.e. ways for Cody to understand codebases. Before joining Sourcegraph, he worked on high-throughput, low-latency ML pipelines at a large scale in adtech. Honors MSc degree in Computer \& Data Science from the University of Ljubljana. Open-source contributor. Research interests include deep learning, neural network embeddings, and model compression.

\begin{acks}
We would like to thank the entire Sourcegraph Cody team for their contributions.
\end{acks}

\bibliographystyle{ACM-Reference-Format}
\bibliography{bibliography}

\end{document}